\documentclass[aps,preprint,showpacs]{revtex4}
\usepackage{graphicx}% Include figure files

\begin{document}

\title{Reverse the force direction at long distance by a running coupling}
\author{Xiang-Song Chen$^{1,2}$%
\footnote{Email: cxs@scu.edu.cn, cxs@chenwang.nju.edu.cn}}
\affiliation{$^1$Department of Physics, Sichuan University,
                Chengdu 610064, China\\
	   $^2$Joint Center for Particle, Nuclear Physics and Cosmology, 
                Nanjing University, Nanjing 210093, China}
\date{\today}
                                                                            
\begin{abstract}
We reveal that a suitable running coupling $\alpha(q^2)$ can reverse the direction of the one-particle-exchange (OPE) force at long distance if (and under general assumptions, only if) the exchanged particle has a mass $m$, either intrinsic or effective. The essential requirement on $\alpha(q^2)$ is an extrapolating restriction $\alpha (-m^2)<0$. The running of $\alpha(q^2)$ can be from either renormalization-group evolution or certain derivatives in the Lagrangian. Reversal of the OPE force direction at long distance (namely, attraction transits to repulsion, or vice versa) may have important implications for gravity and cosmic acceleration, particle and nuclear physics, and also condensed matter properties such as superconductivity. 
\end{abstract}
%11.10.Hi Renormalization group evolution of parameters
%11.10.Jj Asymptotic problems and properties
\maketitle

In this short note we discuss an interesting and fundamental phenomenon that a running coupling does not merely mean the increase or decrease of interaction strength as the energy (or distance) scale varies. Surprisingly, a running coupling may as well alter the attractive-repulsive nature of a force. Certainly the attraction-repulsion transition along the interacting distance is well-known for van der Waals and nuclear forces, which are due to complicated processes. Instead, what we examine here is the simple one-particle-exchange (OPE) force. We hope to open a new avenue towards the understanding of why sometimes Nature exhibits forces contrary to the usual expectations. The most famous examples include the recent discovery of accelerated cosmic expansion \cite{snova} which implies repulsive gravity at cosmological scale, and superconductivity which requires attractive force between electrons.

In the standard approach, the static OPE potential with quantum corrections is given by Fourier transformation of the complete propagator:  
\begin{equation}
V(r)= \frac{\xi}{2\pi^2}\int d^3 q \frac{e^{i\vec q\cdot \vec r}}
{\vec q^2+m^2-\Pi^*(\vec q^2)} \alpha.
\label{normal}
\end{equation}
Here $\xi$ is the sign factor of the interaction; for the electric interaction $\xi=1$ or $-1$ as the two interacting objects have the same or opposite charges, while for the gravitational interaction $\xi\equiv -1$.  
$m$ is the mass of the exchanged particle ($m$ can be zero), $\alpha$ is the renormalized coupling constant, $\Pi^*$ is the complete self-energy from all proper contributions. At tree level, Eq. (\ref{normal}) gives the familiar potential $V_0(r)=\xi \frac{\alpha}{r} e^{-mr}$. 
To include radiative corrections, we propose a most convenient way of 
adopting the off-shell renormalization scheme for {\em each} Fourier 
component. 
In this way the propagator in Eq. (\ref{normal}) is always renormalized 
to be the free form, and all quantum effects manifest now in the 
scale-dependence of the running coupling constant $\alpha(\vec q^2)$:  
\begin{eqnarray}
V(r)&=& \frac{\xi}{2\pi^2}\int d^3 q 
\frac{e^{i\vec q\cdot \vec r}}{\vec q^2+m^2} \alpha(\vec q^2)
\nonumber \\
&=&\frac{\xi}{i\pi r}\int_{-\infty}^{+\infty}dq 
\frac{q e^{iqr}}{q^2+m^2} \alpha(q^2).
\label{master}
\end{eqnarray}

We will prove that at large enough $r$ the attractive-repulsive property 
of $V(r)$ is opposite to that of $V_0(r)$ if the following conditions are met for $\alpha(q^2)$:

\begin{enumerate}
\item $ \alpha(q^2)$ is positive definite, and $\alpha(q^2)/q$ drops 
monotonically to zero faster than $q^{-n}$ ($n>0$) as $q\to \infty$. 

\item $ \alpha(q^2)$ can be extrapolated to be an analytical function 
$\tilde\alpha(z)$ in the upper complex plane with no singularity up to 
$\mathrm{Im}(z)=\lambda>(1+1/n)m$. [In most cases, $\tilde\alpha(z)$ is just
$\alpha (z^2)$.]

\item There exists a however small angle $\phi_0$;
the function $f(z)\equiv \tilde \alpha(z) z/(z^2+m^2)$ satisfies
$|f(qe^{i\phi})|\leq 2|f(q)|$ for $\phi\in [0,\phi_0]\cup[\pi-\phi_0,\pi]$. 

\item $\tilde \alpha(im)<0$, or $\alpha(-m^2)<0$ as 
$\tilde \alpha(z)=\alpha(z^2)$.
\end{enumerate}

In fact, only Condition 4 is essential, the first three are very loose. 
With Condition 2, the integration of $f(z)e^{izr}$ along the wide contour in 
Fig. 1 gives
\begin{equation}
\frac {\xi}{i\pi r} \oint _{ABCDA}dz
\frac{ \tilde \alpha(z) z e^{izr}}{z^2+m^2} = \xi
\frac{\tilde \alpha (im)}{r}e^{-mr}\equiv V_\infty (r).
\label{contour}
\end{equation}

\begin{figure}
\includegraphics[width=12.0cm]{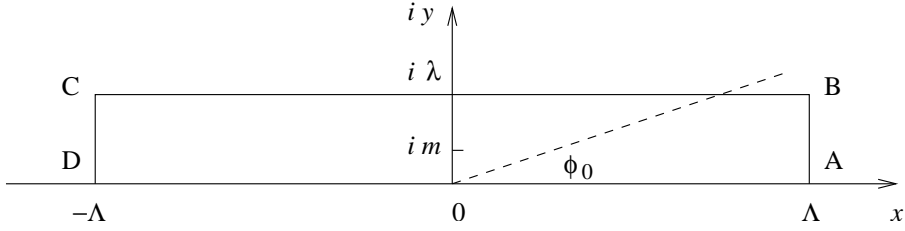}
\caption{Integration contour ABCDA}
\end{figure}

In what follows, we analyze the contour integration part by part,
and prove that as $r\to \infty$ the quantum-corrected OPE potential $V(r)$ 
approaches $V_\infty (r)$, whose attractive-repulsive property is opposite to
that of $V_0(r)$ by Condition 4.

\begin{eqnarray}
I_{DA} &\equiv&\frac {\xi}{i\pi r}\int_{DA}dz f(z)e^{izr}\\
       &=& V(r)-I_{A+}-I_{D-},\\
I_{A+}&\equiv& \frac {\xi}{i\pi r} \int_{\Lambda}^{\infty} dq f(q) e^{iqr},\\
I_{AB}&\equiv& \frac {\xi}{i\pi r} \int_0^{\lambda}idy 
		f(\Lambda+iy)e^{i\Lambda r-yr},\\
I_{BC}&\equiv& \frac {\xi}{i\pi r} \int_\Lambda^{-\Lambda} dq 
		f(q+i\lambda)e^{iqr-\lambda r}.
\end{eqnarray}
Analogous expressions for $I_{D-}$ and $I_{CD}$ are omitted. 

With Condition 1 it can be easily seen by integration by parts that 
\begin{equation}
|I_{A+}| \leq  
%\frac {1}{\pi r^2}\left(f(\Lambda)+\int_\Lambda^{\infty} dq |f'(q)|\right)=
\frac {2}{\pi r^2} f(\Lambda) < \frac {2}{\pi r^2}\Lambda^{-n}.
\end{equation}
Thus $| I_{A+} |$ (and analogously $|I_{D-}|$) can be made much smaller than 
$V_\infty(r)$ if we take  
\begin{equation}\label{lower}
\Lambda \gg e^{mr/n}/|r\tilde \alpha (im)|^{1/n}.
\end{equation}

On the other hand, 
\begin{eqnarray}
\vert I_{AB}\vert &\leq& \frac {1}{\pi r^2} 
	(1-e^{-\lambda r})\mathrm{Max}|f(z)|_{z\in AB}, \\
\vert I_{BC}\vert &\leq& \frac{2}{\pi r} \Lambda
	 e^{-\lambda r} \mathrm{Max}|f(z)|_{z\in BC}.
\end{eqnarray}
Condition 3 tells that $|f(z)|\leq 2 |f(q)|$ for 
$|\mathrm{Re}(z)|=q >\lambda/\phi_0$. 
So, for $\Lambda >\lambda/ \phi_0$ we have 
$\mathrm{Max}|f(z)|_{z\in AB}<2f(\Lambda)<2\Lambda ^{-n}$. Then  
$|I_{AB}|$ (and analogously $|I_{CD}|$) can be 
also made much smaller than $V_\infty (r)$ if $\Lambda \gg 
\mathrm{Max}\{\lambda/\phi_0,e^{mr/n}/|r\tilde \alpha (im)|^{1/n}\}$, 
which can be absorbed into 
Eq. (\ref{lower}) for large enough $r$. 

Finally, since Condition 1 says that $f(q)$ decreases for large $q$,
for $|\mathrm{Re}(z)|=q>\lambda/\phi_0$ we have 
$|f(z)|\leq 2 f(q) <2 f(\lambda/\phi_0)$. Therefore, 
the maximum of $|f(z)|_{z\in BC}$ must be at a 
finite $z$, which we denote as $z_0$. Then we have
\begin{equation}
|I_{BC}|\leq \frac{2}{\pi r} \Lambda e^{-\lambda r} |f(z_0)|.
\end{equation}

For this to be much smaller than $V_\infty(r)$, we need 
\begin{equation}\label{upper}
\Lambda \ll e^{(\lambda -m)r}|\tilde \alpha (im)|/|f(z_0)|. 
\end{equation}

It is now ready to reach the conclusion: since 
$(\lambda-m)>m/n$ by Condition 2, we see that as long as $r$ is 
large enough $\Lambda$ can be chosen to satisfy both Eq. (\ref{lower}) 
and Eq. (\ref{upper}), then as $r\to \infty$ the sum 
$(|I_{A+}|+|I_{D-}|+|I_{AB}|+|I_{BC}|+|I_{CD}|)$ is negligible compared 
to $V_\infty (r)$, and $V(r)$ approaches 
$V_\infty (r)=\xi \tilde \alpha (im)e^{-mr}/r$, as we aimed to prove. 
This asymptotic behavior tells a critical difference
between $m=0$ and $m\neq 0$: In the massless case the OPE potential 
$V(r)$ behaves ``regularly;'' 
at infinity it is essentially not influenced by quantum 
corrections, and always recovers the classical limit $\xi \alpha(q^2=0)/r$. 
In the massive case, however, at infinity $V(r)$ always differs from the 
classical limit, and quantum corrections may even reverse its sign.
Such distinction in asymptotic behaviors is schematically plotted in Fig. 2.

\begin{figure}
\includegraphics[width=12.0cm]{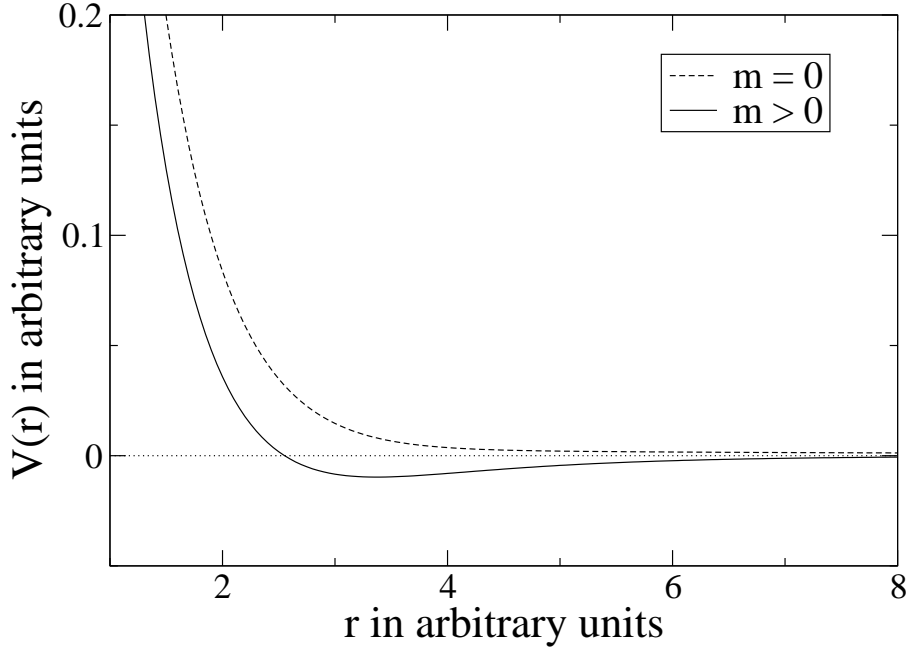}
\caption{Schematic exhibition of different asymptotic behaviors of the  OPE potential $V(r)$. Dashed line: $m=0$; solid line: $m\neq 0$ and $\alpha(q^2)$ satisfies the four conditions listed in the text,
which shows that $V(r)$ transits from repulsive to attractive  
as $r$ increases. The units are arbitrary.}
\end{figure}

In the above discussions we have assumed that the running of $\alpha(q^2)$ is due to renormalization-group effect. However, it is very important to remark that the running can also originate from certain derivatives in the Lagrangian. The most obvious case is derivative coupling to fields, which introduces factor of $q^2$ into $\alpha(q^2)$. On the other hand, if the tree-level propagator of the exchanged particle takes a non-standard form, then any factors other than $1/(q^2+m^2)$ should also be absorbed into the ``running coupling'' in our formulation in Eq. (\ref{normal}). 

Undoubtedly, the long-distance transition of the fundamental one-{\em massive}-particle-exchange force from attractive to repulsive (or vice versa) could have critical implications for physics of many fields. Very recently, we discussed the realization of such phenomenon in fourth-derivative gravity with a tiny graviton mass, which can lead to an ideal unification of dark matter and cosmic acceleration \cite{chen}. It would be  very interesting to conjecture whether such transition could occur in condensed matter and cause crucial change of material properties (say, superconductivity); and as well in particle and nuclear physics where many force-mediating mesons are massive.

This work was supported by China NSF grant 10475057.

\end{document}